\documentclass[prl,nobibnotes,floatfix,superscriptaddress,twocolumn,showpacs]{revtex4-1}
\usepackage{amsfonts,amsmath,graphicx,epsfig,latexsym}
\usepackage{subfigure}
\usepackage{graphicx}
\usepackage{color}
\usepackage{natbib}
\usepackage{multirow}
\usepackage{latexsym,amssymb}
\usepackage{amsmath}

\begin{document}
\title{  Pressure Suppression of Electron Correlation in the Collapsed Tetragonal Phase of $CaFe_2As_2$ : A DFT-DMFT Investigation} 

\author{Subhasish Mandal}
\affiliation
{Geophysical Laboratory, Carnegie Institution of Washington, Washington D.C. 20015, USA}
\author{R. E. Cohen}
\affiliation
{Geophysical Laboratory, Carnegie Institution of Washington, Washington D.C. 20015, USA}
\affiliation
{Department of Earth Sciences, University College London, Gower Street, WC1E 6BT, London, United Kingdom}
\author{K. Haule}
\affiliation
{Department of Physics, Rutgers University, Piscataway, New Jersey 08854, USA}

\begin{abstract}
{\footnotesize

Recent studies reveal a pressure induced transition from a paramagnetic tetragonal phase (T) to a collapsed tetragonal  phase (CT) in $CaFe_2As_2$, which was found to be superconducting  with non-hydrostatic pressure at low temperature.  
We have investigated the effects of electron correlation and local fluctuating moment in both tetragonal and collapsed tetragonal phases of the paramagnetic $CaFe_2As_2$ using self consistent DFT+DMFT  with continuous time quantum Monte Carlo as the impurity solver. From the computed optical conductivity, we find a gain in the optical kinetic energy  due to the loss in Hund's rule coupling energy in the CT.  We find that the transition from T to CT turns $CaFe_2As_2$ from a bad metal into a good metal. 
 Computed mass enhancement and local moments also show significant decrease in the CT, which confirms the suppression of the electron correlation in CT phase of $CaFe_2As_2$. }

\end{abstract}


\maketitle
\newpage

The discovery of superconductivity in Fe-based compounds with $T_c$  in the range from 26 to 56 K has  created a new paradigm in condensed matter physics \cite{doi:10.1021/ja800073m,Mazin:2010he,iron1}. 
The effect of magnetism on the superconducting and normal state properties of unconventional superconductors like  cuprates and Fe-pnictides has gained wide interest with the discovery of antiferromagnetic (AFM) ground state near superconductivity \cite{delaCruz:2008ej,doi:10.1021/ja800073m,Nphys1}. Suppression of the AFM or spin density wave state by doping or pressure was found in various families of Fe-pnictidies \cite{RevModPhys.83.1589}.  Superconductivity in these materials is very sensitive to pressure, and applied pressure has become an important tool to test  different theories and to understand the mechanism of superconductivity, which is still a puzzle.  One of the major questions in high $T_c$ superconductors  is the nature of the magnetism, the strength of the correlation and its role in superconductivity.  Whether magnetism in Fe-based materials arises from weakly correlated itinerant electrons \cite{Mazin:2008jj} or it requires some degree of electron correlations \cite{PhysRevLett.100.226402,Aichhorn:2010dp} and localization of electrons \cite{PhysRevB.78.020501,PhysRevLett.101.076401} is presently a subject of debate \cite{mag1,PhysRevLett.101.076401}. Hence  it is important to know whether the nature of magnetism in Fe-based superconductors requires a description that only takes into account  Fermi surface nesting, the effect of local moment or a combination of both.

 In the Fe-pnictide family, ``122"  compounds with $AFe_2As_2$ (A=Ca, Sr, Ba) are widely studied systems, where $T_c$ can reach as high as 38 K \cite{PhysRevLett.101.107006,RevModPhys.83.1589}. In  the ``122" family, $CaFe_2As_2$ is found to be  unique,  where superconductivity emerges  upon application of modest non-hydrostatic pressure \cite{Torikachvili:2008cy}. With hydrostatic pressure it undergoes a structural transition from a tetragonal phase (T) to a collapsed tetragonal phase (CT)  \cite{PhysRevB.78.184517}. Another study found that superconductivity develops within the collapsed tetragonal phase of $Ca_{0.67}Sr_{0.33}Fe_2As_2$ under pressure \cite{PhysRevB.85.184501}.  The CT phase in $CaFe_2As_2$ is characterized by a $\sim$10 \% reduction in the c-axis of the room temperature tetragonal phase. 
Magnetic and electronic structures are found to be strongly influenced by this transition in both pure and rare-earth doped $CaFe_2As_2$. 
For example, an increase in As-As hybridization due to the suppression of magnetic moment \cite{Yildirim:2009ce}, a 
topological change in the Fermi surface due to Lifshitz transition \cite{Coldea:2009ei}, and  
quenching of Fe local moment in the low temperature CT phase was observed \cite{Gretarsson:2013ic}.  
In addition, disappearance of the AFM order \cite{PhysRevB.85.184501}, suppression of spin fluctuations \cite{PhysRevB.79.060510}, and recovery of Fermi liquid behavior \cite{PhysRevB.83.060505} were also found in the CT phase.   
We ask several questions for Ca122: 1) What is the role of applied pressure in  the CT phase? 2) What is the role of electron correlation for this transition?  3) What are the sizes of the fluctuating local moments in both phases of Ca122?  
  
  Here we try to address these questions by studying optical, magnetic and electronic properties using the combination of density functional theory (DFT) and Dynamical Mean Field Theory (DMFT). 

\begin{figure}
\includegraphics[width=220pt, angle=0]{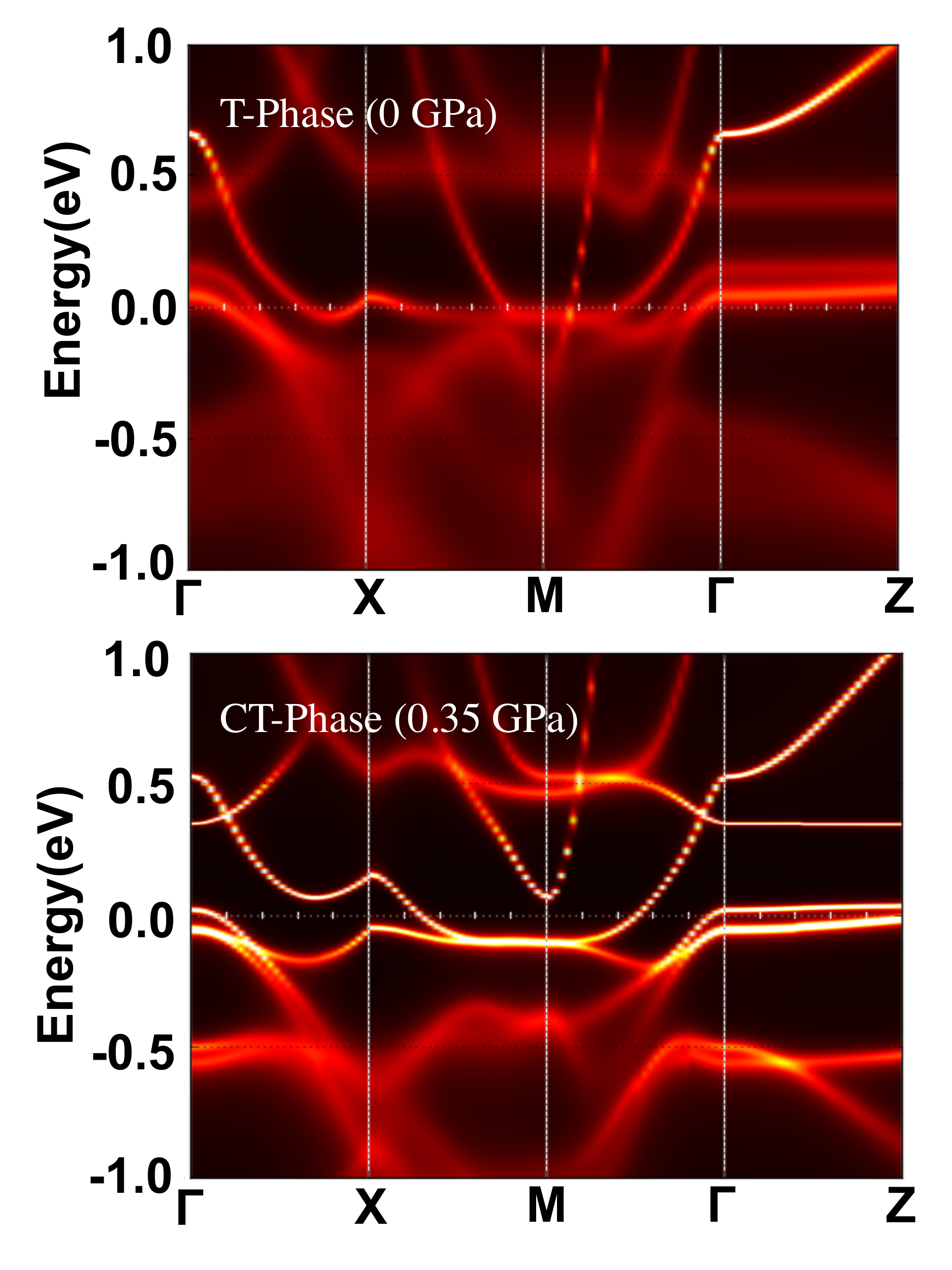}
\caption{ (Color online)
  DFT+DMFT spectral function for (a) T-phase and (b) CT-phase indicating  an  incoherence-coherence crossover for the bands crossing the Fermi energy due to modest applied pressure.  }
\end{figure}

{\it Methods.-} To capture the local moment physics in paramagnetic material like Fe-pnictides,  one needs to go beyond conventional density functional theory (DFT). DFT in combination with dynamical mean field theory (DMFT)  (DFT+DMFT) has proved to be a good approximation to describe  fluctuating local moment and electron correlation \cite{haule_spin,Wang:2013gz}.  The structures and the atom positions used here are taken from the neutron scattering measurement \cite{PhysRevB.78.184517}. In the DFT-DMFT method, the self-energy, sampling all Feynman diagrams local to the Fe ion, is added to the DFT Kohn-Sham Hamiltonian \cite{RevModPhys.78.865,PhysRevB.81.195107}.  This implementation is fully self-consistent and all-electron \cite{PhysRevB.81.195107,haule3}. The computations are converged with respect to charge density, impurity level, chemical potential, self-energy, lattice and impurity Green's functions.  The lattice is represented using the full potential linear augmented plane wave method, implemented in the Wien2k \cite{wien2k} package in its generalized gradient approximation (PBE-GGA). The continuous time quantum Monte Carlo method is used to solve the quantum impurity problem and to obtain the local self-energy due to the correlated Fe {\it 3d} orbitals.
The self-energy is analytically continued from the imaginary to real axis using an auxiliary Green's function.
The Coulomb interaction $U$ and Hund's coupling $J$  are fixed at 5.0 eV and 0.7 eV, respectively \cite{Kutepov:2010bu}. A fine k-point mesh of $10\times 10 \times 10$ and total 80 million Monte Carlo steps for each iteration are used for the paramagnetic phase of the $CaFe_2As_2$ at room temperature.  Here we study electronic and optical properties of $CaFe_2As_2$  in its paramagnetic phase as a function of compression and specially investigated electronic correlation and local moment  in T and CT phases.

 {\it Spectral function.-} We describe  computed orbital resolved spectral function ($A(k,\omega)$) in Fig. 1. We noticed a significant change in the sharpness of the DMFT spectral function for the bands that are close to the Fermi energy ($E_F$). Going from T to CT phases, the DMFT spectral function becomes more coherent. This indicates the suppression of correlation in the CT phase. We found significant changes in the topology of the Fermi surface in the CT phase, similarly predicted by DFT calculations \cite{Coldea:2009ei}. Specially 2D  cylindrical hole bands become flat in the CT phase and the 2D bands that were above the $E_F$ in T-phase are below  $E_F$ in CT-phase.

{\it Optical properties.-}  We computed the in-plane (averaged over x and y directions)  optical conductivity  ($\sigma_1 (\omega)$)  for standard DFT, DFT+DMFT and compared that with experiments performed at ambient pressure with a single crystal of $CaFe_2As_2$  \cite{Nakajima:2010hh}. 
   DFT overestimates the spectral weight for the low energy part of the spectra, but the optical conductivity computed in the DFT+DMFT method agrees well with the experimental optical conductivity (Fig. 2a).  
  To investigate the strength of correlations  and to quantify the reduction of the Drude response compared to band theory in pure Ca122,  we looked at the spectral weight from the real part of the optical conductivity. 
We use a truncated version of the f-sum rule \cite{PhysRevB.72.224517},  similarly to Ref. \cite{Schafgans:2012kl}. The experimental or theoretical optical kinetic energy $K$, which is proportional to the spectral weight of the Drude component of the optical response, can be determined by integrating  the real part of the optical conductivity up to a cutoff frequency $\Omega$:
\begin{equation}
K(\Omega) = \frac{120}{\pi}\int_0^\Omega   \sigma_1 (\omega) d\omega
\end{equation}

 \begin{center}
 \begin{figure*}
\includegraphics[width=440pt, angle=0]{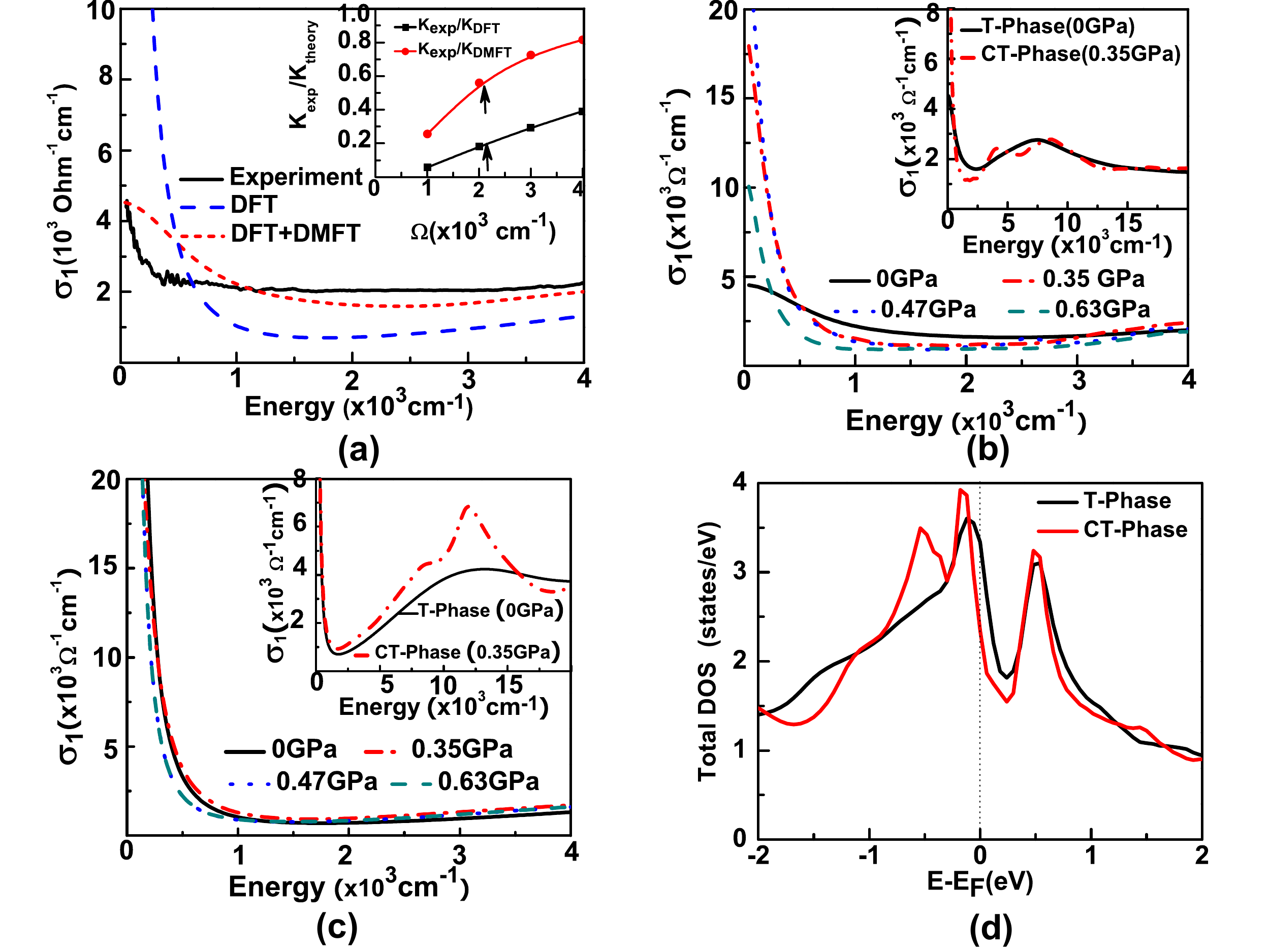}

\caption{(Color online).
Optical conductivity and density of states (DOS) of $CaFe_2 As_2$ in the T and CT phase: (a) Comparison of the real part of the optical conductivity at P=0 between experiment and theory; inset shows the ratio of the experimental and theoretical optical kinetic energy as a function of integration cut off frequency ($\Omega$), the solid arrow represents the possible cut off frequency determined from the minima of $\sigma(\omega)$. Experimental conductivity is reproduced from Ref. \cite{Nakajima:2010hh}.  Calculated in plane average of the optical conductivity as a function of compression with (b) DFT+DMFT and (c) DFT methods; insets show the high energy optical conductivity. (d) DOS calculated in DFT+DMFT for T and CT phases.}
\end{figure*}
\end{center}

\begin{center}
\begin{figure*}
\includegraphics[width=440pt, angle=0]{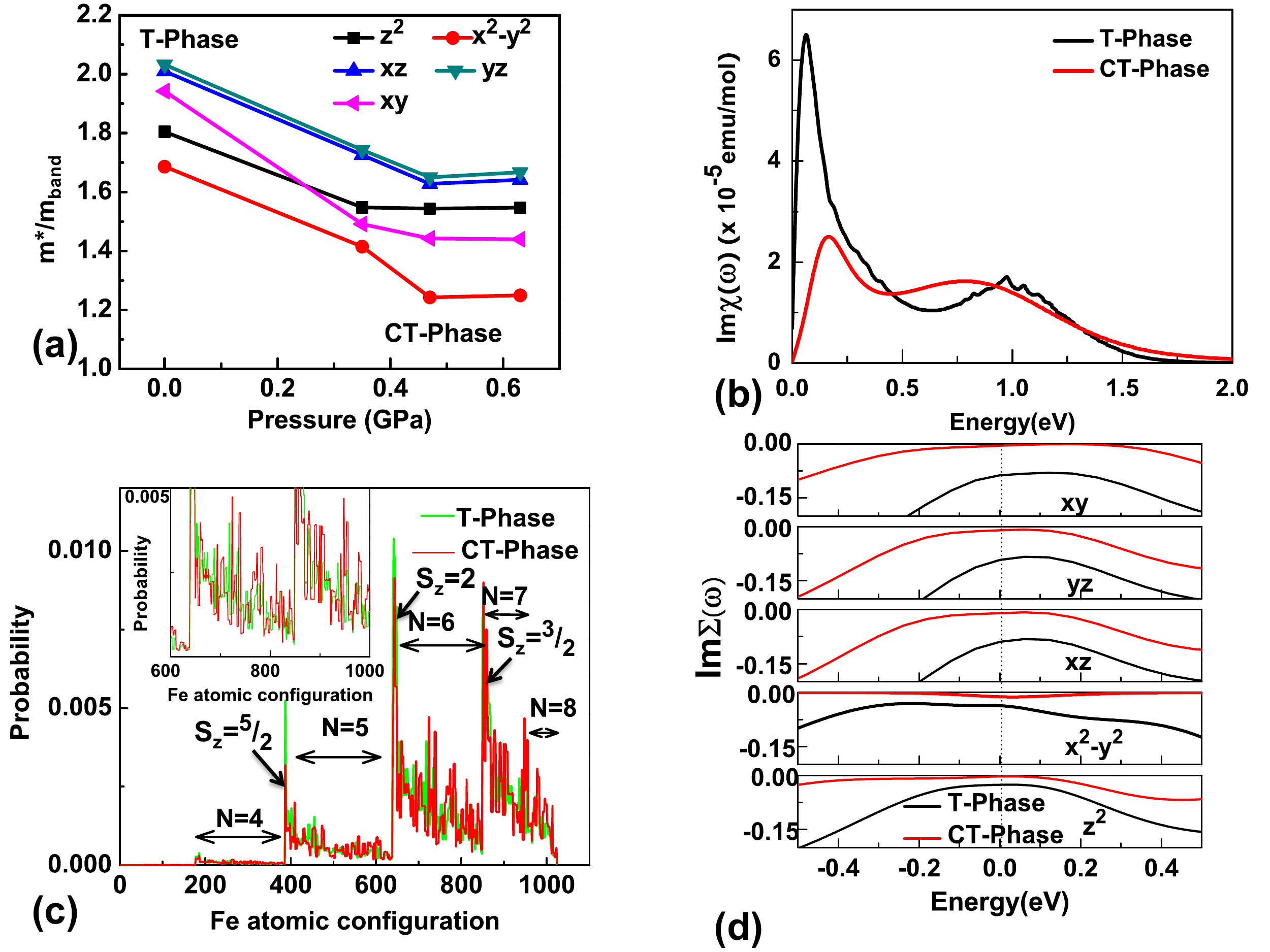}
\caption{(Color online).
DFT+DMFT calculated (a) mass enhancement ($m^*/m_{band}$), (b) Imaginary part of the local dynamic magnetic susceptibility, (c) Atomic histogram of Fe {\it 3d} shell, and (d) Orbital resolved imaginary part of the self-energy for both T and CT phases.}
\end{figure*}
\end{center}

We used the experimental infrared conductivity data from Ref. \cite{Nakajima:2010hh} to calculate the experimental optical kinetic energy at ambient pressure. 
We took a similar approach as in Ref. \cite{Schafgans:2012kl}, where the cutoff value is considered in such a way that it should be high enough to account for all the Drude weight but not so high as to include significant contributions from the inter-band transitions. Similarly we computed $K_{DFT}$ and $K_{DMFT}$, where $K_{DFT}$ and $K_{DMFT}$ are the optical kinetic energies calculated in the DFT and DFT+DMFT methods respectively. 
At ambient pressure we then normalize experimental optical kinetic energy ($K_{exp}$) to $K_{DFT}$. This ratio is often used to describe the degree of electron correlation. For the extremely correlated case of a fully localized Mott insulator like the cuprate parent compounds, $K_{exp}$/$K_{DFT} $ $\sim$  0, whereas in electronically uncorrelated materials such as a fully itinerant metal such as copper, the ratio of  $K_{exp}$/$K_{DFT}$  is approximately 1. 
The many-body effects beyond band theroy, such as dynamical correlationdue to on-site Coulomb repulsion and Hund's rule coupling, renormalize the electronic bandwidth and consequently reduce the optical kinetic energy. Hence the ratio   $K_{exp}$/$K_{DFT} $ characterizes the strength of the correlation in a material. 
 We first describe this ratio for P=0 in the inset of Fig. 2a as a function of the cutoff frequency ($\Omega$). $\Omega$  can be determined from the minima of $\sigma_1 (\omega)$. $K_{exp}$ is obtained from the infrared conductivity data from Ref. \cite{Nakajima:2010hh}. The value of $K_{exp}$ is found to be 13830.15, 23512.82, and 33516.87 $cm^{-2}$  and  $\Omega$ = 2000, 3000, and 4000 $cm^{-1}$ respectively; while  $K_{DFT}$ is found to be 76362.16, 80245.7, and 85602.8 $cm^{-2}$ respectively.      We find  $K_{exp}$/$K_{DFT}$ to be 0.18 - 0.39 in the T-phase. We obtained  $K_{DMFT}$  to be 24680.5, 32394.3, and 41094.8 $cm^{-2}$  and  $\Omega$ = 2000, 3000, and 4000 $cm^{-1}$ respectively and the ratio of $K_{exp}$/$K_{DMFT}$ is found to be 0.56-0.81. The Drude weight agrees better with DFT+DMFT method when we compare with a recent experiment \cite{PhysRevB.86.134503} performed at 300K.  A similar value was obtained for Ba122 in paramagnetic sate \cite{Schafgans:2012kl}.   The ratio of the optical kinetic energy becomes larger with larger $\Omega$ as noticed from the inset of Fig. 2a. Therefore, we reconfirm that DFT+DMFT has the ability to accurately describe the optical response in the paramagnetic state.   This also indicates the presence of electron correlation for P=0 in the T-phase of Ca122. 
 
We plot  $\sigma_1 (\omega)$  as a function of pressure (P)  in Fig. 2b. We see a large spectral weight transfer in the DFT-DMFT method going from T to  CT phase within the infrared region, indicating the increase in electron's kinetic energy. Going from T to CT upon application of pressure, Ca122 changes from a bad metal to a good metal. This  transition is not seen in DFT  $\sigma_1 (\omega)$. The  $\sigma_1 (\omega)$ calculated in DFT as a function of pressure is almost constant in the infrared region.  Only at higher energy did we notice a peak in the CT-phase (inset Fig. 2c).

To examine in more detail, we compute the spectral weight or electron kinetic energy by using formula (1) for the CT-phase. For a cutoff frequency of 2000 $cm^{-1}$, we found that optical kinetic energy increases from 24680.57 to 33341.34 $(cm^{-1})^2$ in the CT phase (at 0.35 GPa), whereas in DFT it decreases from 76363.43 to 47075.74 $(cm^{-1})^2$.  We then took the ratio of the spectral weight calculated in the DFT-DMFT and DFT  approaches. The ratio of $K_{DMFT}$/$K_{DFT}$ is 0.324 at P=0 and  0.708 at P=0.35 GPa.  A similar trend is found when we take the cut off frequency as 1000 and 3000  $cm^{-1}$. This indicates the suppression of correlations in the CT-phase.  

  We plotted the density of states (DOS) in Fig. 2d. The DOS near $E_F$ decreases in the CT phase when we compare it with the T-phase. So, we argue that the increase in $K_{DMFT}$ in CT is not due to the density of states near $E_F$,  but due to Hund's rule coupling.  Comparing  the histograms, which describe the probabilities of different Fe configurations in solids, we see that  the high-spin states become more probable in the T phase. Thus, local Fe moment is larger for the T-phase  with the enhanced Hund's rule coupling due to their larger lattice constants.

{\it Mass enhancement.-} To further investigate the degree of correlation,
we computed  mass-enhancement ($m^*/m_{band}$) = $1/Z_A$, where $Z_A$=$(1-\frac{\delta \Sigma}{\delta \omega})^{-1}_{\omega=0}$. In a Fermi-liquid  $Z_A$  is the quasiparticle weight, which is unity for a non-interacting system, and 
is much smaller than unity for a strongly correlated system.
We have calculated $m^*/m_{band}$ for all the Fe-d orbitals and plotted them as a function of P in Fig. 3a.
Going from T-phase to CT-phase, we notice a drop in the $m^*/m_{band}$ for all $d$-orbitals. 
First, we noticed that  the $d_{z^2},d_{x^2-y^2}$  orbitals are less correlated and the $t_{2g}$ orbitals ($d_{xz} $, $d_{yz}$, and $d_{xy}$) are more correlated at P=0. 
With increasing pressure, electron correlation becomes weaker for all $d$-orbitals. Especially the effect of pressure on $m^*/m_{band}$ is mostly dramatic on the $d_{xy}$ orbital. For example,  calculated $m^*/m_{band}$ is 2.01 for $d_{xy}$ orbital at P=0 GPa  and 1.63 at P=0.47 GPa.  In the CT-phase $m^*/m_{band}$ almost remains same with increasing P.

{\it Local dynamical magnetic susceptibility.-}
To infer the effect of pressure on the fluctuating magnetic moments, we compute the dynamic magnetic susceptibility, which measures the spatial and temporal distribution of the magnetic fluctuations.  In Fig. 3b we plot Im[$\chi(\omega)$] on real frequency for both T and CT phases. The continuos  time quantum Monte Carlo impurity solver is used to obtain the local dynamic susceptibility $\chi(\imath\omega )$ as a function of Matsubara frequencies. We analytically continued the data using maximum entropy method to obtain Im[$\chi(\omega)$] on real frequency. We notice a sharp peak in $\chi(\omega )$ at low energy ($\sim$ 0.19 eV) indicating large fluctuating moment \cite{PhysRevB.89.125113}, which is very pronounced in the T-phase. The peak height decreases in the CT, reflecting a substantial reduction in local moment and hence confirms that the fluctuating local moment is reduced in the CT phase.

{\it Hund's rule interaction.-} The iron pnictides/chalcogenides are considered to be Hund's metals \cite{Yin:2011ca,haule3}. Instead of  the Hubbard interaction (U), the Hund's rule interaction causes the quasiparticle mass enhancement in these materials \cite{Yin:2011ca,haule3}.  Electrons with the same spin but different orbital quantum numbers are aligned by the Hund's rule interaction when they find themselves on the same iron atom. DFT+DMFT method can truly capture the Hund's rule physics. To quantify the probability of finding an iron atom in the solid in one of the atomic states, we present the atomic histogram for both T and CT in Fig. 3c.  The DMFT atomic basis is constructed from the five 3{\it d} orbitals of an iron atom, that  spans a Hilbert space of size $2^{10}$ = 1024 for 10 different occupancies with N=0, 1,...10. Here the first (last) few states with a particular N show the high (low) spin state.  In Fig. 3c we clearly see the spikes in probability for the high spin states  at the beginning of the constant N interval. As a consequence, the low spin states, at the end of the constant N interval, lose substantial weight. In the absence of Hund's coupling, the high and the low spin states would be equally probable.  From Fig. 3c. we notice that in the CT  state the high-spin states become less probable and the low spin states become more probable (inset of Fig. 3c).  This shows an overall loss of the Hund's rule coupling energy in CT due to reduced lattice constant. As a consequence, the low-energy part of the self-energy (Fig. 3d) shows a clear change in Im$[\Sigma(\omega)]$ in the CT. 
 
In summary, we have computed the correlated electronic structure for $CaFe_2As_2$ for ambient pressure tetragonal phase and high pressure collapsed tetragonal phase. We found a significant gain in the electronic kinetic energy in the CT phase due to the loss of the Hund's coupling energy.  Increasing optical kinetic energy  reflects the suppressions of electron correlation in the CT. Our results are consistent with a recent NMR study where suppression of electron correlation was found in the low temperature CT phase \cite{PhysRevB.89.121109}. Computed mass enhancement and the paramagnetic fluctuating moment also reflects the suppression of the electron correlation.

 We thank V. Struzhkin for helpful discussions and M. Nakajima for sending us the experimental data on optical conductivity.  We thank Jane Robb for helping us in editing the manuscript. This research was supported as part of EFree, an Energy Frontier Research
Center funded by the US Department of Energy Office of Science, Office of Basic Energy Sciences under Award DE-SC0001057 and Carnegie Institution of Washington. K. H acknowledges the supports from NSF DMR 0746395. REC is supported by the Carnegie Institution and by the European Research Council advanced grant ToMCaT. Computations were performed at the NERSC supercomputing facility.

\bibliography{super}
\end{document}